\documentclass[twocolumn,showpacs,showkeys,superscriptaddress]{revtex4-2}
\usepackage{amsmath}
\usepackage{amsfonts}
\usepackage{amssymb}
\usepackage{graphicx}

\begin{document}
	
	\title{Link auxiliary field method in the Extended Hubbard model}
	\author{Sergey Mostovoy}
	\email{sd.mostovoy@physics.msu.ru}
	\affiliation{Faculty of Physics, Moscow State University, Moscow, 119991, Russia}
	\affiliation{National Research Center ``Kurchatov Institute'', Moscow, 117218, Russia}
	
	\author{Oleg Pavlovsky}
	\email{pavlovsky@physics.msu.ru}
	\affiliation{Faculty of Physics, Moscow State University, Moscow, 119991, Russia}
	\affiliation{National Research Center ``Kurchatov Institute'', Moscow, 117218, Russia}
	
	\begin{abstract}
	The aim of this work is to generalize the method of Hubbard fields in fermion Quantum Monte Carlo simulation to the case of link fields. The introduced Hubbard link fields play a role of the interaction fields responsible for the attraction and repulsion of electronic excitations at the nodes. Such improvements leads to a number of advantages concerning thermalization time, autocorrelations, metastable states and probability distributions of the quantities. It is demonstrated that computations using five fields are more stable and reliable in case of energy observables. \vspace{2pt}
	\end{abstract}

	
		\maketitle
	
		
		
	

	\section{Introduction}
	
	The Monte Carlo method of modeling quantum systems has become one of the most important tools in condensed matter physics, nuclear physics and high-energy physics. The method has shown to be extremely effective in solving strong coupling problems and for studying systems with long correlations. Nowadays the Monte Carlo method turns to be most popular approach available to study a large quantum system directly from the fundamental principles of quantum theory.
	
	The study of electronic properties of graphene, as well as other semimetals, is an example of such challenges. Graphene quasiparticles are massless\cite{PhysRev.71.622, katsnelson_2012}, so graphene physics becomes extremely non-local. A large effective coupling constant (with the electric field) makes it also a strongly interacting theory. Thus, the physics of electronic excitations of electron states in graphene is an important area where the Monte Carlo methods are applied.
	
	A number of important results in graphene physics have been obtained by the Monte Carlo method: the possibility of forming an antiferromagnetic condensate \cite{buividovich2012numerical,ulybyshev2013monte}, the influence of defects \cite{valgushev2013influence,ulybyshev2015magnetism} and magnetic field \cite{braguta2013numerical,braguta2013numerical2,boyda2014numerical}, the Casimir effect \cite{buividovich2017interelectron}. However, a lot of difficulties remain while applying the Monte Carlo method to fermion systems. One of the most important is the \textit{sign problem} \cite{PhysRevD.102.054502}, which prevents Monte Carlo methods to be used in the important case of a nonzero chemical potential. Despite great efforts, a simple, universial and effective method to overcome the sign problem has not yet been found.
	
	Statistical physics describes the change of systems's states through the concept of a \textit{configuration space}. In case of graphene (described by the Extended Hubbard model's Hamiltonian) this configuration space reveals a very complex landscape which represents another significant difficulty in connection with the Monte Carlo method. It is shown\cite{PhysRevD.101.014508} that the configuration space of the extended Hubbard model on a hexagonal lattice is a collection of valleys with a relatively low value of action $S$. The valleys are separated by high domain walls. A Hybrid Monte Carlo process cannot leave such a valley during a finite time period after entering it. As a result, a simulation observes only some of valleys, but not the entire configuration space. To solve this problem, it is proposed \cite{PhysRevB.98.235129} to introduce additional Hubbard fields in order to increase the number of degrees of freedom. In addition to the traditional Hubbard field associated with on-site interaction of charges, it was proposed to introduce a Hubbard field associated with the interaction of spins. It has been shown that such an enhancement significantly facilitates Monte Carlo simulations in the extended Hubbard model.
	
	In the paper, it is shown that the original idea \cite{PhysRevB.98.235129} can be developed further. It is proposed to consider auxiliary fields attached to \textit{links}, and not just to nodes. In the context of BSS QMC \cite{BSS} this technique was used, for example, in \cite{ModPhysLetB}. Thus, the net number of Hubbard fields in a model on a hexagonal lattice can be increased from two to five. What goals this complication serves for? First of all, this increases dimensions of the configuration space and makes it easier for the Monte Carlo process to bypass domain walls \cite{PhysRevE.106.025318} in the expanded space of Hubbard fields. This helps to reduce temperatures at which the Monte Carlo simulation is efficient and relatively easy. Secondly, the study of complex observables associated with a large number of creation and annihilation operators becomes more accessible because a degree of divergences in statistical distribution is reduced when additional field variables are added. In the paper, we deal with energy of the electron excitations and their energy squared which contain 4, 6 or 8 operators multiplied.
	
	The paper is organized as follows. In the Section 2 the formulas defining the model analytically are given and the procedure for introducing link fields is shown. Section 3 demonstrates the results of the technical tests of a computer program, which prove a correctness of simulations and the enhancement in the performance. In Section 4 energy-like observables are calculated within a wide range of temperatures in order to compare their features in case of 2- and 5-fields approaches. Finally, the Conclusion summarizes the results obtained and outlines the plans for further research.
	
	\section{Model description}
	\label{sec:model}
	
\begin{figure}
	\centering
	\includegraphics[width=0.35\linewidth]{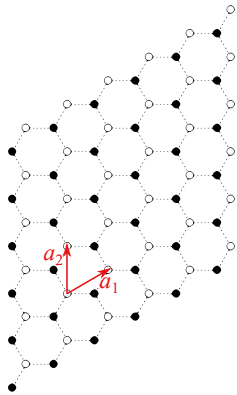}
	\caption{A geometry of the lattice used in computations. Two primitive vectors of the lattice are shown, periodicity along which is imposed.}
	\label{fig:bc}
\end{figure}
	
	Let us consider an extended Hubbard model on a hexagonal lattice. A plane of $2 L^2$ sites is enclosed with periodic boundary conditions of Born-von K\'{a}rm\'{a}n type as presented in \figurename~\ref{fig:bc}. The Hamiltonian reads:
	\begin{equation}\label{eq:hamil}
		\hat{H} = -\kappa\sum\limits_{\langle x,y \rangle,\sigma } \left( \hat{a}^\dagger_{x,\sigma} \hat{a}_{y,\sigma} + \mbox{h.c.} \right) + \frac12 \sum\limits_{x,y} V_{xy} \hat{q}_x \hat{q}_y,
	\end{equation}
	where $\kappa$ is the hopping parameter, $\langle x,y \rangle$ indicates adjacent sites, $\sigma$ stands for a spin projection (up or down), $\hat{q}_x = \hat{a}^\dagger_{x,\uparrow}\hat{a}_{x,\uparrow} + \hat{a}^\dagger_{x,\downarrow}\hat{a}_{x,\downarrow} - 1$ defines an electric charge on a site, with $\hat{a}_{x,\sigma}$ being an annihilation operator for $\sigma$-spin particle at node $x$. $V_{xy}$ refers to a matrix of electrostatic interaction with parameters $V_{00}$ and $V_{01}$ depending on a distance between two electrons (on-site and nearest-neighbours are involved). The dimensions of the matrix are $N_s \times N_s$, where $N_s = 2L^2$. The coefficients are chosen to be $\kappa = 2.8$ (eV), $V_{00} = 3.5\kappa$, $V_{01} = 0.8\kappa$. This set corresponds to semi-metal --- a complex state with a complicated configuration landscape. The calculation itself is performed by the Quantum Hybrid Monte Carlo method. A relevant technical description can be found in \cite{PhysRevB.89.195429}, and the procedure for discretizing the partition function within a path integral approach is described in details in \cite{PhysRevB.98.235129}. The total number of time slices amounts to $2N_T$, because each of $N_T$ transitions in Euclidean time is quantized into two factors, concerning kinetic and interaction part of the Hamiltonian. $N_T = 60$ is used throughout the computations. Our computer code can work in a mode using two Hubbard fields and computations implement the following partition function ($\beta = 1/T, k_B=1$):
	\begin{equation}\label{eq:partfnc}
		\begin{split}
		\mathcal{Z} &= \mathrm{Tr} \left\{ e^{-\beta \hat{H}} \right\} = \\ &\int\mathcal{D}[\zeta,\zeta^*]\mathcal{D}[\phi,\chi] \exp \left\{-S_{\mbox{HS}} - \zeta^*(M M^\dagger)^{-1}\zeta \right\},
	\end{split}
	\end{equation}
	the action of Hubbard fields is ($\delta={\beta}/{N_T}$)

	\begin{equation}\label{eq:actionHS}
		S_{\mbox{HS}} = \frac{\delta^n}{2} \phi \tilde{V}^{-1} \phi + \frac{\left(\chi - (1-\alpha)V_{00} \delta^{1-m} \right)^2}{2(1-\alpha)V_{00}} \delta^n,
	\end{equation}
	\begin{equation}\label{eq:intmatrix}
		\tilde{V}_{xy} = \alpha V_{00} \delta_{xy} + (1-\delta_{xy}) V_{xy},
	\end{equation}
	where fermionic matrix represents
	
	\begin{equation}\label{eq:fermmatrix}
		M = 
		\begin{bmatrix}
			1 			& 0		  	& 0			& 0			& 0	 & \dots 	& -E^{(f)}_{N_T-1} \\
			E^{(k)}_0 	& 1 	  	& 0			& 0			& 0	 &  		& 0 \\
			0 			& E^{(f)}_0 & 1 		& 0			& 0  &  		& 0 \\
			0 			& 0 	  	& E^{(k)}_1 & 1 		& 0	 &  		& 0 \\
			0 			& 0 	  	& 0 		& E^{(f)}_1 & 1	 &  		& 0 \\
			\vdots 		& 	  		& 			&   		&  	 & \ddots 	& \vdots \\
			0 			& 0		  	& 0			& 0			& 0  & \dots 	& 1
		\end{bmatrix},
	\end{equation}
	\begin{equation}\label{eq:matrixexp1}
		E^{(k)}_t = -\exp \{\kappa \delta \sum\limits_{ \substack{x,y, \\ \langle a,b \rangle } } \left( \delta_{ax}\delta_{by} +  \delta_{ay}\delta_{bx} \right)\},
	\end{equation}
	\begin{equation}\label{eq:matrixexp2}
		E^{(f)}_t = -\mbox{diag} \exp \{ -\delta^m \left( i\varphi_{xt} + \chi_{xt} \right) \},
	\end{equation}
	where $E^{(f)}_t$ are diagonal $N_s \times N_s$ matrices with space indices $x, y$, $E^{(k)}_t$ are dense $N_s \times N_s$ matrices, $\alpha = 0.95$ is a mixing parameter for Hubbard fields of $\varphi_{xt}$ and $\chi_{xt}$, $t = \overline{0,N_T-1}$ serves as a reference number of a time slice (with total number of $2N_T$). In order to derive (\ref{eq:actionHS}) one should use the following form of the Hubbard-Stratonovich transform, with an additional parameter $n$ which helps to control time-discretization errors ($n$ and $m$ are tied by $2m-n=1$):
	
	\begin{equation}\label{eq:HStransform}
		\exp \left\{ -\frac12 \hat{q} V \hat{q} \delta \right\} = \int \mathcal{D}[\varphi] \exp \left\{ -\frac{\delta^n}{2} \varphi V^{-1} \varphi -i \delta^m \varphi \hat{q} \right\}.
	\end{equation}

	In the current paper, $n=-1$ will be fixed. However, a dependence of Hubbard fields' amplitudes $n$ is an interesting question itself. We have launched our program for a set of $n$ and have derived to a conclusion that $n$ should be chosen to be negative, however all values from the list $-1$, $-2$, $-3$ are good enough, so the difference in results and stochastic properties of Monte Carlo time series can be a topic of our further investigation. Of course, $n$ and $m$ are not limited to integers, so this choice is a matter of taste only.
	
	
	The principal \textit{innovation} in our work is an introduction of the auxiliary \textit{link} fields that supplement the main $\varphi_{xt}$ and $\chi_{xt}$ fields defined on the nodes. It is assumed that these additional degrees of freedom greatly increase the dimension of a configuration space, which, firstly, simplifies a generation of a sequence of system configurations during the Monte Carlo flow, and, secondly, reduces the divergences in observables consisting of a large number of creation-annihilation operators. The second circumstance turns out to be significant due to a complicated structure of observables (products of 6 and 8 operators) involved in our research. The problem is related to fermion matrix zeros in the configuration space of the model because a propagator corresponds to elements of the inverse fermionic matrix. The Monte Carlo flow depends on the zeros which can disturb a numeric algorithm to walk around configurations. This can impose limits on the selection of configurations. A detailed discussion can be found in \cite{Ulybyshev2017PathIR}.
	
	Let us show how supplementary \textit{link} fields are proposed to be introduced into the model.
	It is suggested that one can isolate a perfect square for the product of operators at distinct points (this is possible due to the commutation property $[\hat{q}_x, \hat{q}_y] = 0$):
	
	\begin{equation*}
	\begin{split}
	& \frac{V_{00}}{2} \sum\limits_x \hat{q}_x^2 + \frac{V_{01}}{2} \sum\limits_{x \neq y} \hat{q}_x \hat{q}_y = \\
	& \frac{V_{00}}{2} \sum\limits_x \hat{q}_x^2 + \frac{V_{01}}{4} \sum\limits_{x \neq y} \left( (\hat{q}_x+\hat{q}_y)^2 -\hat{q}_x^2 - \hat{q}_y^2 \right) = \\
	&\frac{V_{00}}{2} \sum\limits_x \hat{q}_x^2 - 3\frac{V_{01}}{2} \sum\limits_x \hat{q}_x^2 + \frac{V_{01}}{4} \sum\limits_{x \neq y} (\hat{q}_x+\hat{q}_y)^2 = \\
	& \frac12 \sum\limits_x V'_{00}\hat{q}_x^2 + \frac{V_{01}}{4} \sum\limits_{x \neq y} (\hat{q}_x+\hat{q}_y)^2,
	\end{split}
	\end{equation*}

	where $V'_{00} = V_{00} - 3V_{01}$ indicates a modification of a site interaction. Within the bounds of \cite{PhysRevB.98.235129} a site field will be decomposed into two components $\varphi_{xt}$ and $\chi_{xt}$ in the following manner:
	
	$$
	\alpha \frac{V'_{00}}{2} \sum\limits_x \hat{q}_x^2 - (1-\alpha) \frac{V'_{00}}{2} \sum\limits_x \hat{s}_x^2 +
	(1-\alpha) V'_{00} \sum\limits_x \hat{s}_x,
	$$
	where $\hat{s}_x = \hat{a}^\dagger_{x,\uparrow}\hat{a}_{x,\uparrow} - \hat{a}^\dagger_{x,\downarrow}\hat{a}_{x,\downarrow} + 1$ defines spin at a site. The procedure how the expressions of $\hat{q}_x^2$ and $\hat{s}_x^2$ can be rewritten is described in the papers mentioned above, so we pay attention to the link part only:
	
	\begin{equation*}
	\begin{split}
	&\exp \left\{ -\delta\frac{V_{01}}{4} \sum\limits_x (\hat{q}_x + \hat{q}_{x+\mu})^2 \right\} =\\ &\int\mathcal{D}[\xi^{(\mu)}] \exp \left\{ -\frac{{\xi^{(\mu)}}^2}{V_{01}} \delta^n -i \delta^m \sum\limits_x \xi^{(\mu)}_x(\hat{q}_x + \hat{q}_{x+\mu}) \right\}=\\
	&\int\mathcal{D}[\xi^{(\mu)}] \exp \left\{ -\frac{{\xi^{(\mu)}}^2}{V_{01}} \delta^n -i \delta^m \sum\limits_x (\xi^{(\mu)}_x + \xi^{(\mu)}_{x-\mu})\hat{q}_x \right\},
	\end{split}
	\end{equation*}
	where $\mu = 1,2,3$ and $\xi^{(\mu)}$ is a vector composed of $\xi^{(\mu)}_x$.  In the second sum of the last expression a substitution $x+\mu \to x$ is made in order to get $\hat{q}_{x+\mu}$ in the form of $\hat{q}_{x}$. A symbol of $x-\mu$ designates a site in the direction opposite to those of $\mu$ (i.e., $y = x-\mu$ means that a neighbour of $y$ in the direction $\mu$ is $x$). One can show that $\xi$ fields are manifested in the fermion matrix in the following way:
	
	\begin{widetext}
	\begin{equation}\label{eq:matrixexp3}
		E^{(1)}_t = -\mbox{diag} \exp \left\{ -\delta^m \left( \chi_{xt} + i\left( \varphi_{xt} + \sum\limits_\mu \xi^{(\mu)}_{xt} + \sum\limits_\mu \xi^{(\mu)}_{x-\mu,t} \right)  \right) \right\}.
	\end{equation}
	\end{widetext}
	
\begin{figure}[tb]
	\centering
	\includegraphics[width=0.5\linewidth]{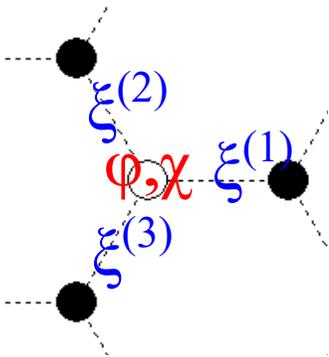}
	\caption{Five fields used in our research: two site fields $\varphi_{xt}$ and $\chi_{xt}$ on each site $x$ together with link fields $\xi^{(\mu)}_{xt}, \mu=\overline{1,3}$ attached to links.}
	\label{fig:latfields}
	\end{figure}

	
	So, we have demonstrated how the dimension of the configuration space of the model can be increased to 5 (see \figurename~\ref{fig:latfields}). In this work, an attempt is made to compare the results of such ``expended'' calculations with standard ones in cases of order parameters and energetic quantities, so that an inaccuracy in values obtained can be reduced. One should expect to observe the difference between results obtained with 2 and 5 fields because additional degrees of freedom may make Monte Carlo configurations more chaotic and independent. Moreover, features of observables' distributions \cite{PhysRevE.106.025318} in the configuration space are affected. The data presented below will show that this is the case. The second opportunity, selection of $n$, manages the amplitude of the fields. The consequences of this will be studied in our subsequent work.
	
	One can note, that $V'_{00} = V_{00} - 3V_{01}$ has appeared in the formulae. The fact that a set of points $(V_{00}, V_{00}/3)$ on a phase diagram of the model represent the boundary of the area available to this method of calculations is known \cite{PhysRevB.89.205128}, but in case of two fields is ``hidden'' in the structure of eigenvalues of $\tilde{V}^{-1}$. In our case (five fields) it becomes explicit and obvious. This can be regarded as a beautiful touch of the 5-field formalism.
	
	It is well-known that a long-range potential plays some non-trivial role in graphene \cite{PhysRevB.99.205434}. It is worth saying that the method how supplementary link fields are introduced does not preclude including further interaction radii into $V_{xy}$ such as $V_{02}$ or $V_{03}$ \cite{Phys.Rev.Lett.106}.
	
	\section{Technical properties}
	
	In order to verify how supplementary link fields influence a configuration space landscape one can observe an evolution of $\varphi_{xt}$, $\chi_{xt}$ and $\xi^{(\mu)}_{xt}$ during a thermalization process. After some program launches were performed, it turned out that significantly less algorithm steps are needed to reach a region of well-thermalized configurations (this was tested by analyzing moving averages of fields, the action $S$ and the autocorrelation times of several observables) when using 5 fields in comparison with a 2-field case. A moving average of $M$ successive computations of an observable $x$ during a Monte-Carlo series is defined as follows:
	$$
	\frac{1}{M} \sum_{i=1}^{M} x_i.
	$$
	In the course of the work, the measurement number $M$ required to stabilize this value was estimated.
	
\begin{figure}
	\centering
	\includegraphics[width=0.9\linewidth]{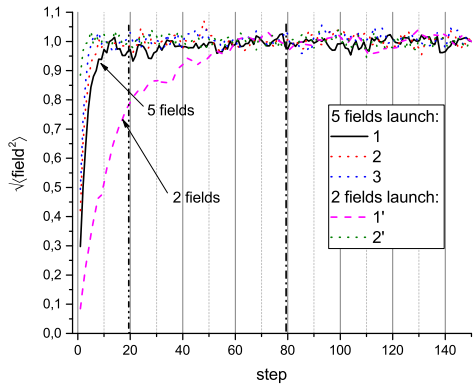}
	\caption{An example of the fields' evolution obtained in two program launches for lattice $6 \times 6$. Mean-square fields are shown. Labels $1$, $2$ and $3$ refer to $\varphi_{xt}$, $\chi_{xt}$ and $\xi^{(2)}_{xt}$ for a 5-field launch while labels $1'$ and $2'$ mean $\varphi_{xt}$ and $\chi_{xt}$ for 2-field mode. Values are normalized by a value at step 150. It took 4 times longer for 2-field launch to reach representative configuration states in comparison to a 5-field case (the marks are shown as dash-dot and dash-dot-dot respectively).}
	\label{fig:therm_proc}
\end{figure}

	An example of a simulation program launch is shown in \figurename~\ref{fig:therm_proc}. Thus, most likely, the technical gain in thermalization itself is achieved by reducing the number of steps required to obtain ``typical'' field values for given model parameters. Moreover, a 5-field approach does not require to find a reverse of the interaction matrix $V_{xy}$ which also simplifies coding and debugging (the operation occurs 4 times in the course of the algorithm). The calculation of additional fields (involves the molecular dynamics evolution and the action $S$) does not increase the work time by more than 4.4\%. Taking into account that the number of algorithm steps is reduced by 10-20 times (due to the need for fewer configurations: to obtain a comparable quality of statistics, it became possible to take 1500 measurements instead of 10000; moreover, one can use every generated configuration without skipping a few), it is possible to obtain a significant benefit when obtaining a series of statistical data to determine the average values of the observables.
	
	An important conclusion shall be drawn: a comparison of the thermalization curves shows that the introduction of 5 fields significantly reduces the thermalization time by several times what increases the effectiveness and performance of the Hybrid Monte Carlo.
	
\begin{figure}
	\centering
	\includegraphics[width=0.9\linewidth]{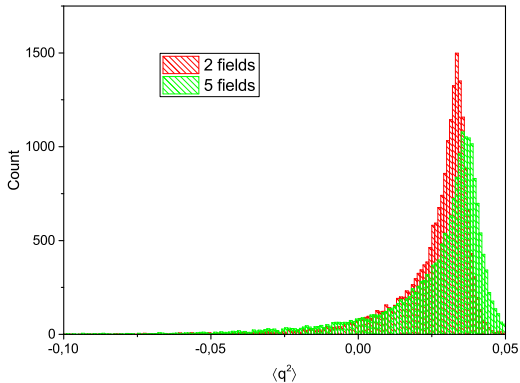}
	\caption{A distribution of $\langle q^2 \rangle$ (per sublattice). It is suggested that in the case of 5 fields a Monte Carlo process involves more distinct values of the observable, so it covers a wider region of a configuration space. The peak is more smooth and slopes grow more gentle. A fitting problem of the probability distribution is discussed in Section \ref{sec:trials}. Computation for $6 \times 6$ and $T = 0.3 \kappa$. Sample size is 20000.}
	\label{fig:hist_q2}
\end{figure}

	The fluctuations sharing between link fields (according to their mathematical definition) seem to be another positive factor. This can help the Markov chain Monte Carlo to observe a configuration space in a flexible manner while generating new configurations. It can be visualized by collecting sequential values of a quantity during a Monte Carlo simulation in order to build a histogram. If one performs this concerning mean of charge squared per sublattice
	\begin{equation}\label{eq:q2oper}
	\langle q^2 \rangle = \frac12 \left\{ \frac{1}{L^4} \left\langle \left(\sum\limits_{x\in A} \hat{q}_x \right)^2 \right\rangle + \frac{1}{L^4} \left\langle \left(\sum\limits_{x\in B} \hat{q}_x \right)^2 \right\rangle \right\},
	\end{equation}
	the result argues that the distribution falls smoother in case of 5 fields and that a larger range of values are presented in the vicinity of the modal (most probable) value (see \figurename~\ref{fig:hist_q2}). The curve itself is reduced in height and more resembles a Gaussian when 5 fields are used. The expectation values remain unchanged though. A long so called heavy-tail is manifested in both the cases, but a numerical comparison of a goodness of the distribution form reveals a difference between 2- and 5-fields results and is discussed for all the observables under the study in the next Section.
	
	A technical feature that differs the usage of two and five fields should be mentioned. It turned out, that the integration step $\Delta\tau$ of the molecular dynamics (an essential part of Hybrid Monte Carlo that obtains ``new'' configurations) varies with temperature in case of five fields if one wants to keep acceptance rate as high as 95-98\%. Specifically, in range of $0.06 < T/\kappa < 0.1$ $\Delta\tau = 0.009$ was used with both 2 and 5 fields while it required $0.008$ at $T = 0.3\kappa$ and $0.005$ at $T = 0.6\kappa$ with 5 fields. Actually, this tuning did not cause a problem at all, but noticeably more independent configurations are obtained which becomes an advantage of the 5-field approach. This can be proved by calculating so-called \textit{autocorrelation times} \cite{doi:10.1142/9789814417891_0003} for Monte Carlo time series. The results of one such investigation are the following (for energy-like observables, for $\langle q^2 \rangle$ the times are less, but still exceed the best values for 5 fields by $2.1$--$2.8$ times).
	
	\begin{table}[h]
	\centering
	\begin{tabular}{|c|c|c|}\hline
	$T/\kappa$ & 2 fields & 5 fields \\\hline
	0.1 & 5.6 & 0.8 \\
	0.15 & 5.2 & 1.2 \\
	0.2 & 4.6 & 1.0 \\
	0.25 & 4.0 & 0.9 \\
	0.3 & 3.6 & 0.8 \\
	0.4 & 2.1 & 1.0 \\
	0.5 & 1.8 & 1.1 \\
	0.6 & 1.5 & 1.2 \\
	0.7 & 1.7 & 1.0 \\
	0.8 & 1.6 & 1.1 \\\hline
	\end{tabular}
	\caption{Autocorrelation times as a function of temperature in cases of 2- and 5-fields approaches.}
	\label{tbl:acc}
	\end{table}

	An important circumstance is the advantage of the 5-field method in a physically interesting temperature range, because it is necessary to perform calculations at low temperatures. It can be expected that the advantage will remain at lower temperatures due to the small temperature dependence of the data for 5 fields, evident from the table. Besides, one can see that in the 2-fields approach, the temperature dependence is significant, so it takes more effort to maintain acceptable statistics while using 2 fields in comparison with 5-fields technic.

	The times can be reduced by increasing $\Delta\tau$, a step of the molecular dynamics, but this has a limit because the acceptance rate lowers. Another approach is to skip several intermediate configurations between sequential ``measurements'', but the strategy is less time-efficient. Thus, it is argued, that 5 fields can help to produce Monte Carlo series of a better quality.

	\section{Trials of energy components}
	\label{sec:trials}
	
	Let us present the results of further tests in favor of Monte Carlo simulations using three additional Hubbard fields $\xi^{(\mu)}_{xt}$. As mentioned above, one of the consequences of expanding the configuration space is a modification of distributions of observables. To check how well this works a number of physical quantities in the form of creation-annihilation operator combinations were computed and statistics of 20000 numbers was investigated. One of meaningful physical quantities with a large number of operators multiplied is the square of the Hamiltonian $\langle \hat{H}^2 \rangle$ which contains products of $8$, $6$, $4$ and $2$ operators. Exact expression of $\langle \hat{H}^2 \rangle$ is outlined in Appendix \ref{appA}.
	
\begin{figure}
		\centering
		\includegraphics[width=0.9\linewidth]{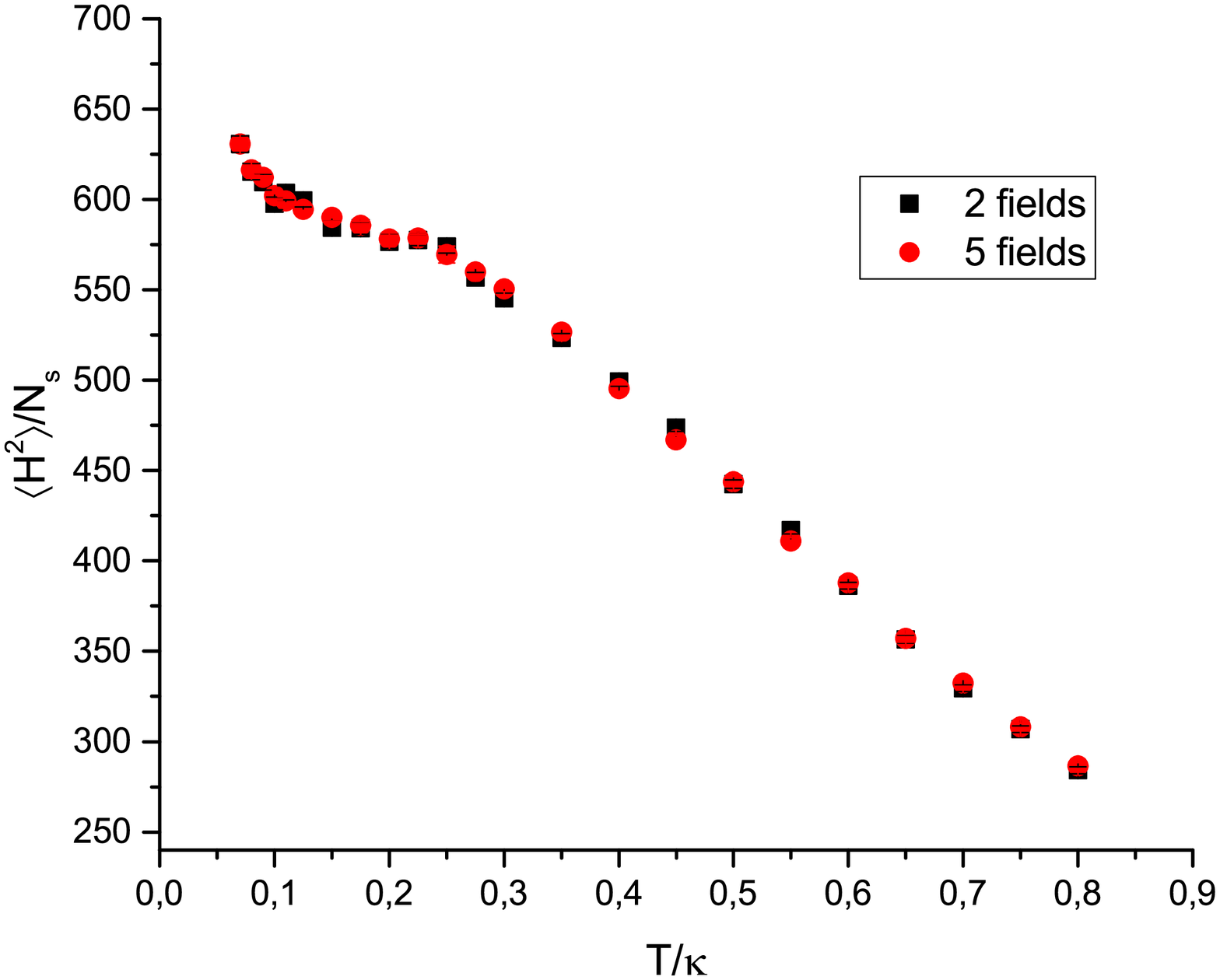}
		\caption{$\langle \hat{H}^2 \rangle$ (per site) for lattice $6\times 6$ as a decreasing function of temperature. $6 \times 6$ lattice. This feature arises due to a contribution of last two terms in (\ref{eq:Hsqr}).}
		\label{fig:H2_temp}
	\end{figure}

\begin{figure}
		\centering
		\includegraphics[width=0.9\linewidth]{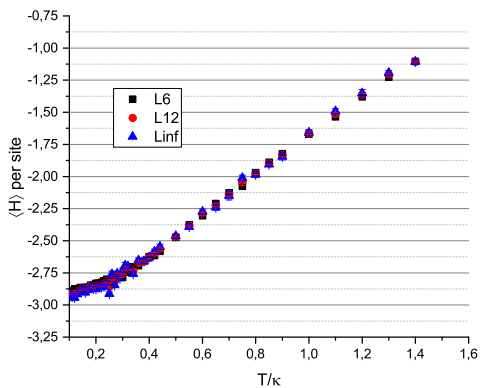}
		\caption{A ``volume effect'' test. Mean energy is computed for a large temperature range for $L=6$ and $L=12$. An infinite volume limit is shown also (labeled as ``Linf''). One can see that the difference is negligible within statistical errors.}
		\label{fig:energy_voleff}
	\end{figure}

	To begin with, one should respect an operator character of the $\langle \hat{H} \rangle$ and compute the Hamiltonian squared in a proper way. Indeed,
	\begin{equation}\label{eq:Hsqr}
	\begin{split}
	\langle \hat{H}^2 \rangle &= \langle (\hat{T}+\hat{U})^2 \rangle = \langle \hat{T}^2 \rangle + \langle \hat{U}^2 \rangle + \langle \hat{T}\hat{U} \rangle + \langle \hat{U}\hat{T} \rangle \\
	& \neq \langle \hat{T} + \hat{U} \rangle^2 = \left(\langle \hat{T} \rangle + \langle \hat{U} \rangle\right)^2
	\end{split}
	\end{equation}
	and $\langle \hat{T}\hat{U} \rangle + \langle \hat{U}\hat{T} \rangle \neq 2\langle \hat{T} \rangle \langle \hat{U} \rangle$. One consequence of (\ref{eq:Hsqr}) is a \textit{descending} behaviour of $\langle \hat{H}^2 \rangle$ with temperature rising (see \figurename~\ref{fig:H2_temp}). One can treat $\langle \hat{T}\hat{U} \rangle$ as $\sum_m \langle \hat{T} |m\rangle \langle m| \hat{U} \rangle$ where $\{ |m\rangle \}$ is a complete set of states. Simulations show that the multipliers have different signs, so this item \textit{reduces} the sum composing $\langle \hat{H}^2 \rangle$. There is no doubt that the square of the energy $\langle \hat{H} \rangle$ itself is an increasing function of temperature, as an energy does (see \figurename~\ref{fig:energy_voleff}).
	
	Secondly, in numerical computations on the lattice, a ``volume effect'' should be checked to be sure of the correctness of the results. This means that when the size of the problem increases, the magnitude of the quantities obtained should not change significantly. Let us demonstrate that this is the case in our calculations of energy observables. \figurename~\ref{fig:energy_voleff} presents computation results of $\langle \hat{H} \rangle$ (per site) in case of $6\times 6$, $12 \times 12$ lattices and extrapolations (point-wise) to the infinite lattice limit. A negligible volume effect with respect to energy quantities can be assured. Note, that such observables as $\langle S^2_z \rangle$ ($S_z$ is the z-component of a spin, the expression has a form of (\ref{eq:q2oper}) with
	$$
	\hat{S}_{z,x} = \frac12(\hat{a}^\dagger_{x,\uparrow}\hat{a}_{x,\uparrow} - \hat{a}^\dagger_{x,\downarrow}\hat{a}_{x,\downarrow})
	$$
	substituted) and $\langle q^2 \rangle$ suffer from severe volume effect (see, for example, \cite{PhysRevB.98.235129}). Now, one can conclude that it is safe to perform energy computations with $L=6$ which requires significantly less CPU time.
	
	This is why we continue by presenting results for $L=6$ ($L=12$ was checked at a few temperatures, of course: volume effect proved to be small enough).
	
\begin{figure}
		\centering
		\includegraphics[width=0.9\linewidth]{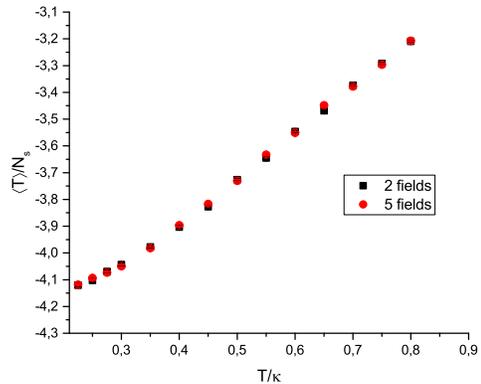}
		\caption{Comparison of kinetic term thermal behaviour for 2-fields and 5-fields approaches. As $\langle \hat{T} \rangle$ has the simplest structure possible, everything looks reliable.}
		\label{fig:kin_cmp}
	\end{figure}

	The results for kinetic term of the Hamiltonian are presented in \figurename~\ref{fig:kin_cmp}. The initial expression reads
	$$
	\langle \hat{T} \rangle = -\kappa \sum\limits_{\langle x,y \rangle,\sigma} \left\langle\hat{a}^\dagger_{x,\sigma} \hat{a}_{y,\sigma} + \hat{a}^\dagger_{y,\sigma} \hat{a}_{x,\sigma} \right\rangle
	$$
	and contains products of 2 operators only, so no mathematical difficulties arise. The graph confirms that everything goes well.
	
	\begin{figure}
		\centering
		\includegraphics[width=0.5\linewidth]{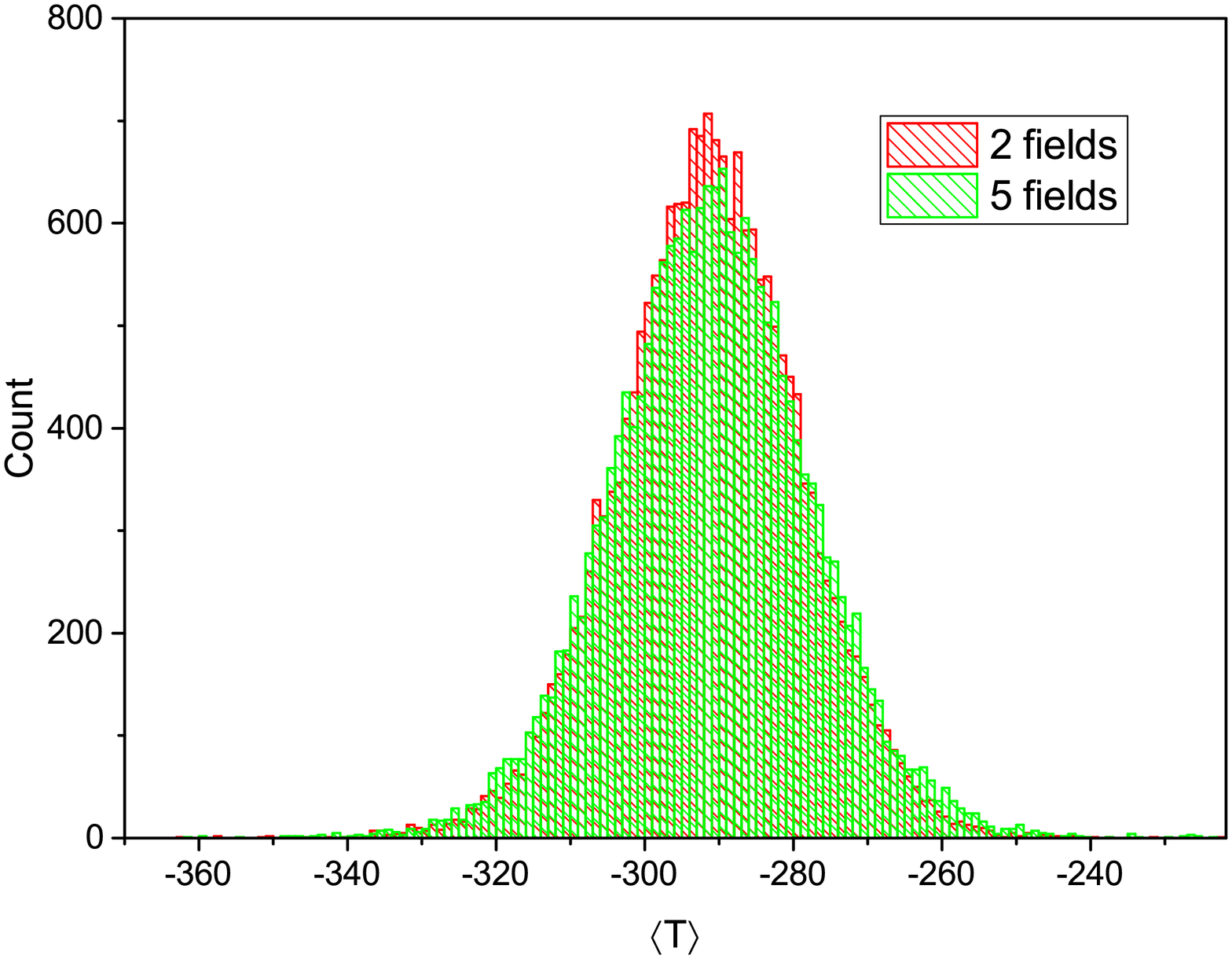}\hfill\includegraphics[width=0.5\linewidth]{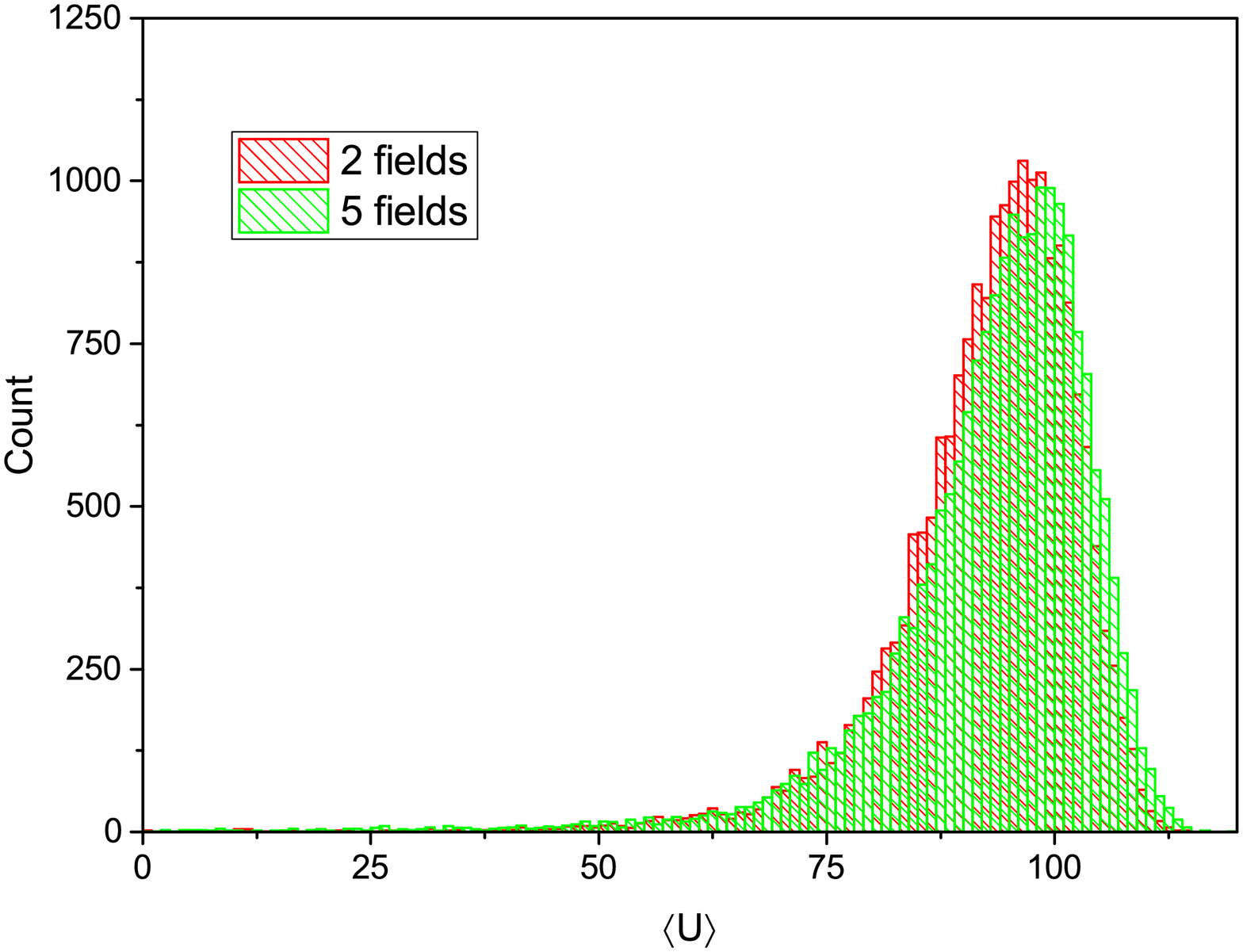}
		\caption{The histograms help to compare features of kinetic and potential energies from a point of view of values presented in distributions obtained for $6 \times 6$. One can see that while $\langle \hat{T} \rangle$ is obvious of a Gauss form, $\langle \hat{U} \rangle$ reveals exponential suppression on the right, but has a long tail in the opposite direction. The latter can cause problems with higher distribution function moments, so it is desirable that a contribution of the tail is reduced. Five fields are helpful in this.}
		\label{fig:hist_T_U}
	\end{figure}

	\noindent Next, one turns to a potential term:
	\begin{equation*}
	\langle \hat{U} \rangle = \langle \hat{U}_1 + \hat{U}_2 \rangle = \frac12 \sum\limits_{x,y} V_{xy} \left\langle \hat{q}_x  \hat{q}_y \right\rangle
	\end{equation*}
	containing products of 4 operators. $U_1$ includes on-site interactions of electrons at the same site ($x=y$) while $U_2$ designates nearest-neighbour interaction. From this point on, some problems arise. \figurename~\ref{fig:hist_T_U} presents the difference between statistical distributions of $\langle \hat{T} \rangle$ and $\langle \hat{U} \rangle$ for a launch of 20000 observables' computations. One can see that configurations with increased interaction energy are depressed exponentially while there are configurations with moderate and weak values of $\langle \hat{U} \rangle$. Nothing of the kind takes place as to $\langle \hat{T} \rangle$. Almost a perfect Gaussian form occurs in the latter case. $\langle \hat{U} \rangle$ reveals a so called heavy-tailed distribution when one of its slopes falls slower than an exponent does. As a result an attempt to compute higher distributions moments such as a variance can run into some difficulties \cite{PhysRevE.106.025318}. So, let us inspect if 5-fields approach can help to enhance observables' distributions form.
	
	\begin{figure}
		\centering
		\includegraphics[width=0.8\linewidth]{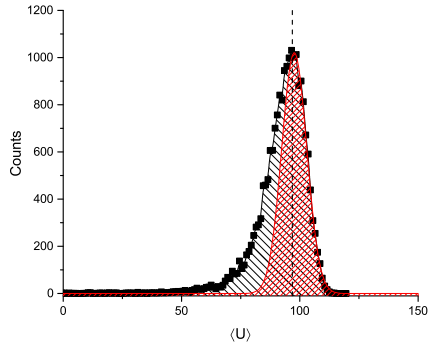}
		\caption{A visualization of the method proposed to compare contributions of a heavy-tail to the net probabilities. A red curve and crossed hatch is the Gaussian fitting of the right side of the data (black points). The single hatch fills an area under the curve of the real distribution. Regions to the left of a dashed line are used to find a fraction of non-Gaussian contribution. An example of 2 fields, $6 \times 6$ lattice. See a description in the main text.}
		\label{fig:hist_U_area}
	\end{figure}

	It is proposed to compare the contributions of heavy-tails as follows. In \figurename~\ref{fig:hist_T_U} the right-hand side of the $\langle \hat{U} \rangle$ distribution resembles a Gaussian, and a fitting confirms this hypothesis. So, one can imagine a perfect symmetrical distribution with the center at point where real distribution maximum exists but of a Gaussian width obtained by fitting the right-hand side. For our example a relative statistical error of the width evaluation is around 3\% only. Next, one computes numerically the area $A$ restricted by the heavy-tailed side (in our case it is the left-hand one) and subtracts the corresponding area of the Gaussian curve. Finally, divide the value by $A$ to estimate how much of the distribution exceeds the desired (Gaussian) curve (see \figurename~\ref{fig:hist_U_area}). In the case of $\langle \hat{U} \rangle$ the quantity computed equals to 39.48\% for 2 fields and 33.68\% for 5 fields. So, the tendency is rather encouraging.
	
\begin{figure}
		\centering
		\includegraphics[width=0.7\linewidth]{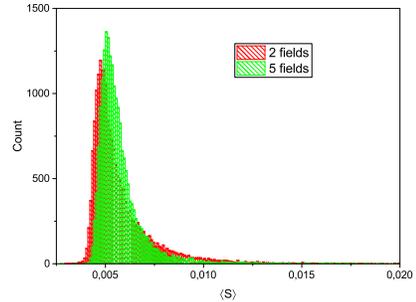}
		\caption{A histogram of values of $\langle S_z^2\rangle$. In the case of 5 fields the tail is obviously lower.}
		\label{fig:hist_S2}
	\end{figure}

\begin{figure}
		\centering
		\includegraphics[width=0.5\linewidth]{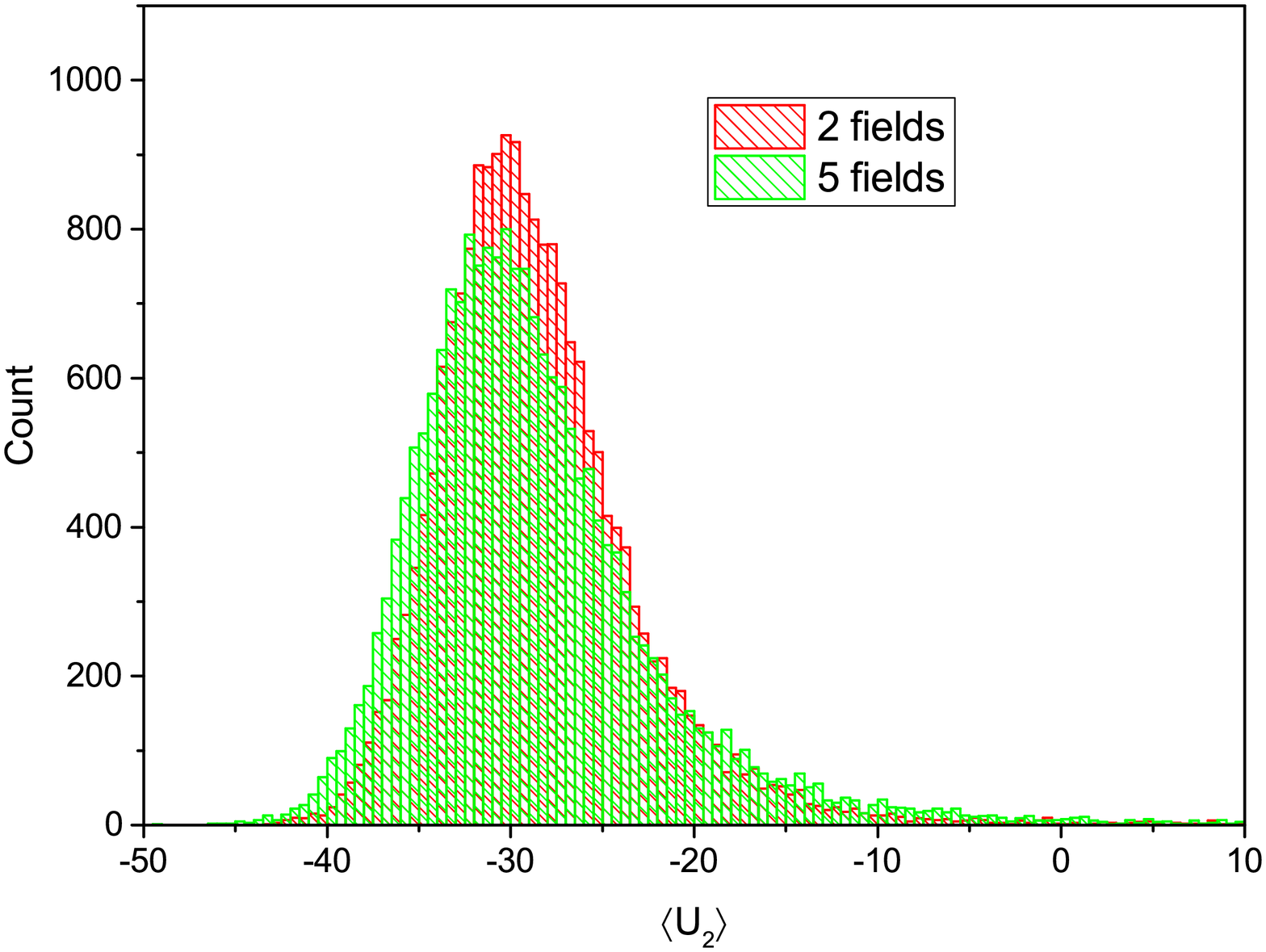}\hfill\includegraphics[width=0.5\linewidth]{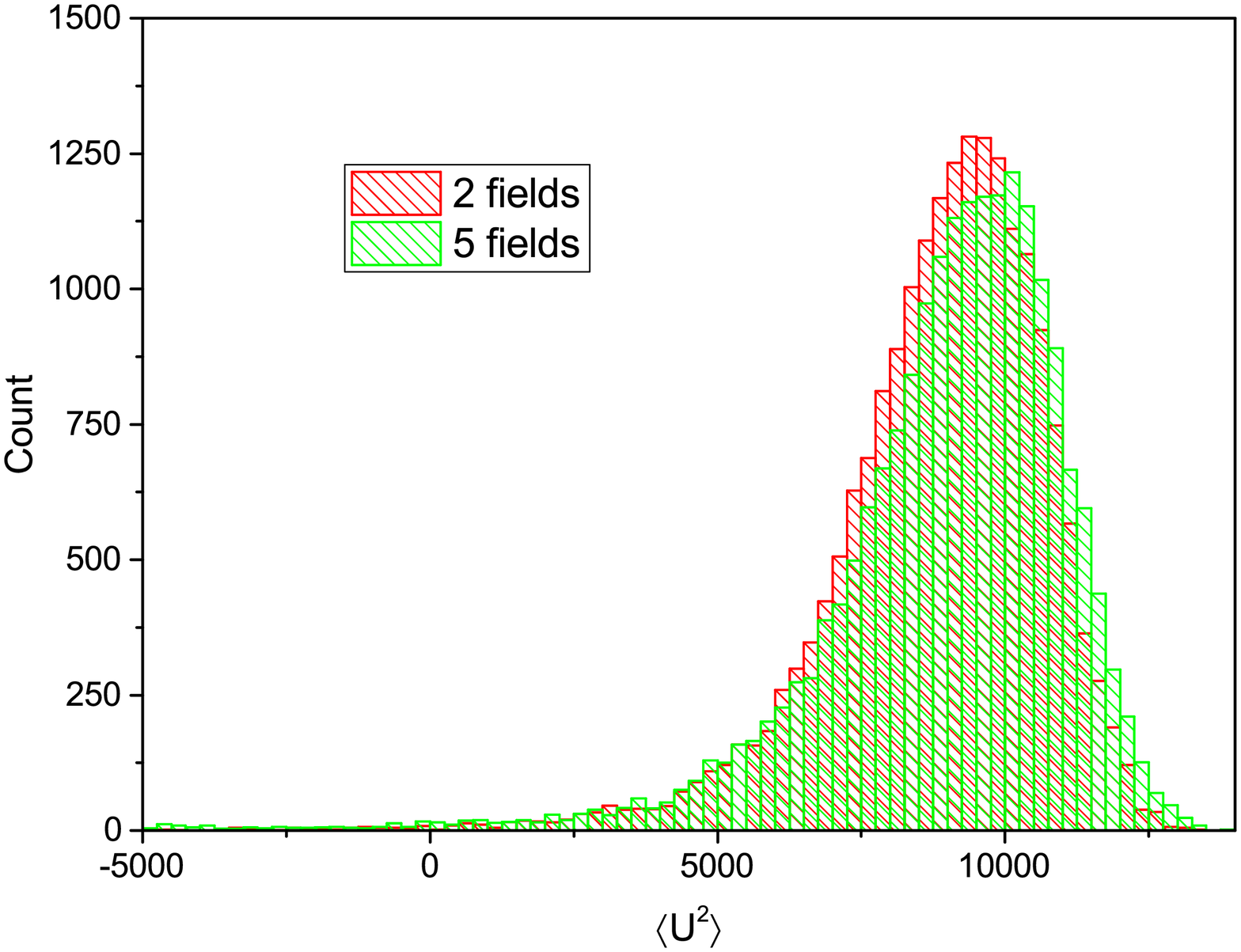}
		\caption{A histogram of values of $\langle \hat{U}_2\rangle$ and $\langle \hat{U}^2\rangle$. The 5-fields result is more like a Gaussian and calculations yield that a long tail is somewhat better in this case.}
		\label{fig:hist_U2_UU}
	\end{figure}

	Actually, for all the observables from the list --- $\langle S_z^2\rangle$, $\langle q^2\rangle$, $\langle T\rangle$, $\langle U\rangle$, $\langle U_1\rangle$, $\langle U_2\rangle$, $\langle \hat{H}\rangle$, $\langle \hat{H}^2\rangle$ --- an area excess of the distribution over the Gaussian (found in the way mentioned above) is reduced when using 5 fields in comparison to a 2-fields approach. The most prominent example is of $\langle q^2\rangle$: 40.7\% for 2 fields and 26.6\% for 5 fields. The examples are presented in \figurename~\ref{fig:hist_q2}, \figurename~\ref{fig:hist_S2} and \figurename~\ref{fig:hist_U2_UU}. The most complex combinations of multiple (up to 8) creation-annihilation operators considered in the paper ($\langle\hat{U}\hat{U}\rangle$) are also improved. The change for $\langle\hat{U}\hat{U}\rangle$ is from 39.19\% to 36.46\%.
	
	\begin{figure}
		\centering
		\includegraphics[width=0.9\linewidth]{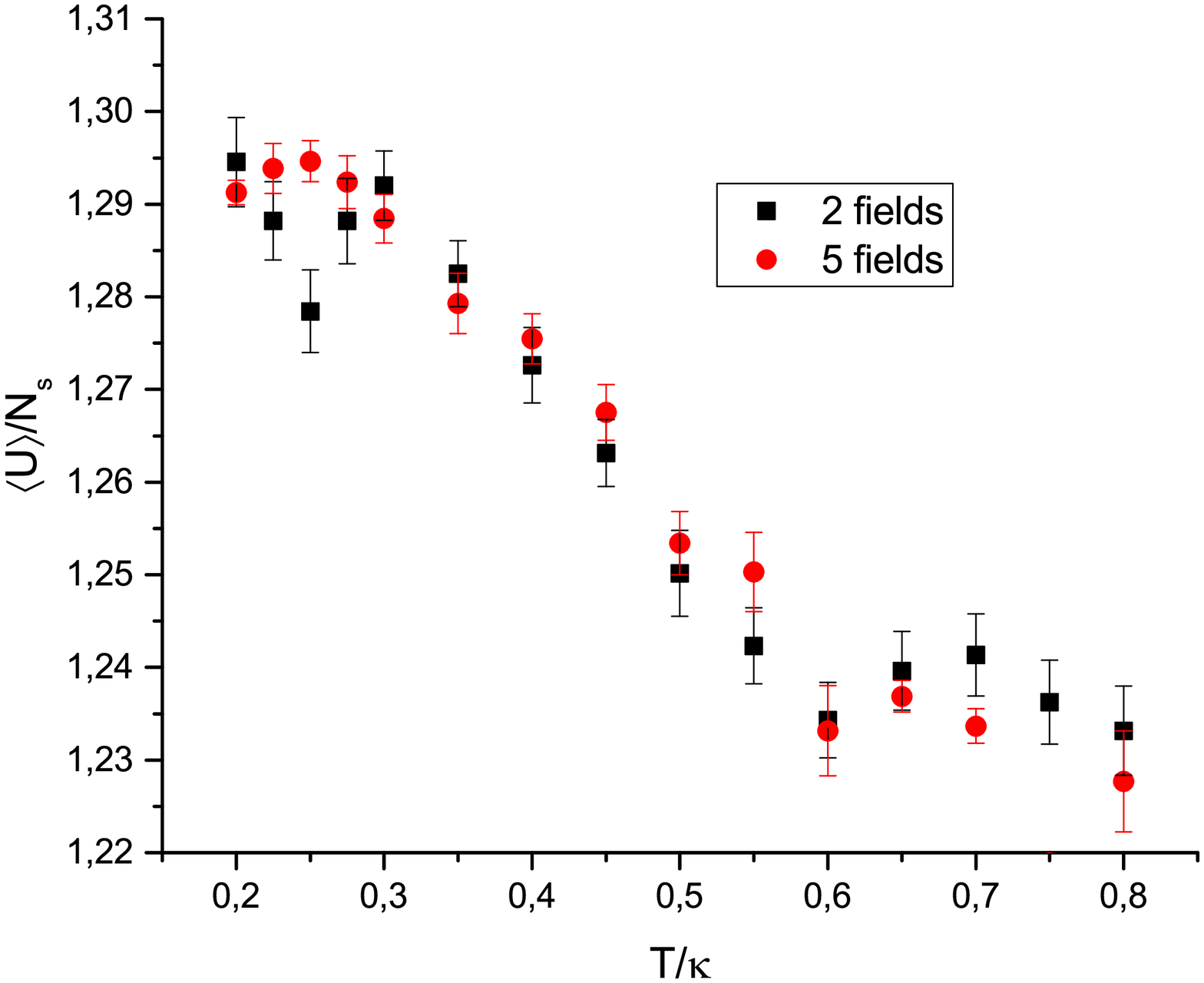}
		\caption{A comparison of the interaction term thermal behaviour for 2-fields and 5-fields approaches. Indications to metastable states are noticeable in case of 2 fields at $0.2 \lesssim T/\kappa\lesssim 0.3$. An observed noise is permitted due to a complicated physical character of a semi-metal state of the Hubbard model in the vicinity of $U=3.2$, $V=0.8$. 5-field approach still yields a more stable result.}
		\label{fig:pot_cmp}
	\end{figure}
	
\begin{figure}
		\centering
		\includegraphics[width=0.9\linewidth]{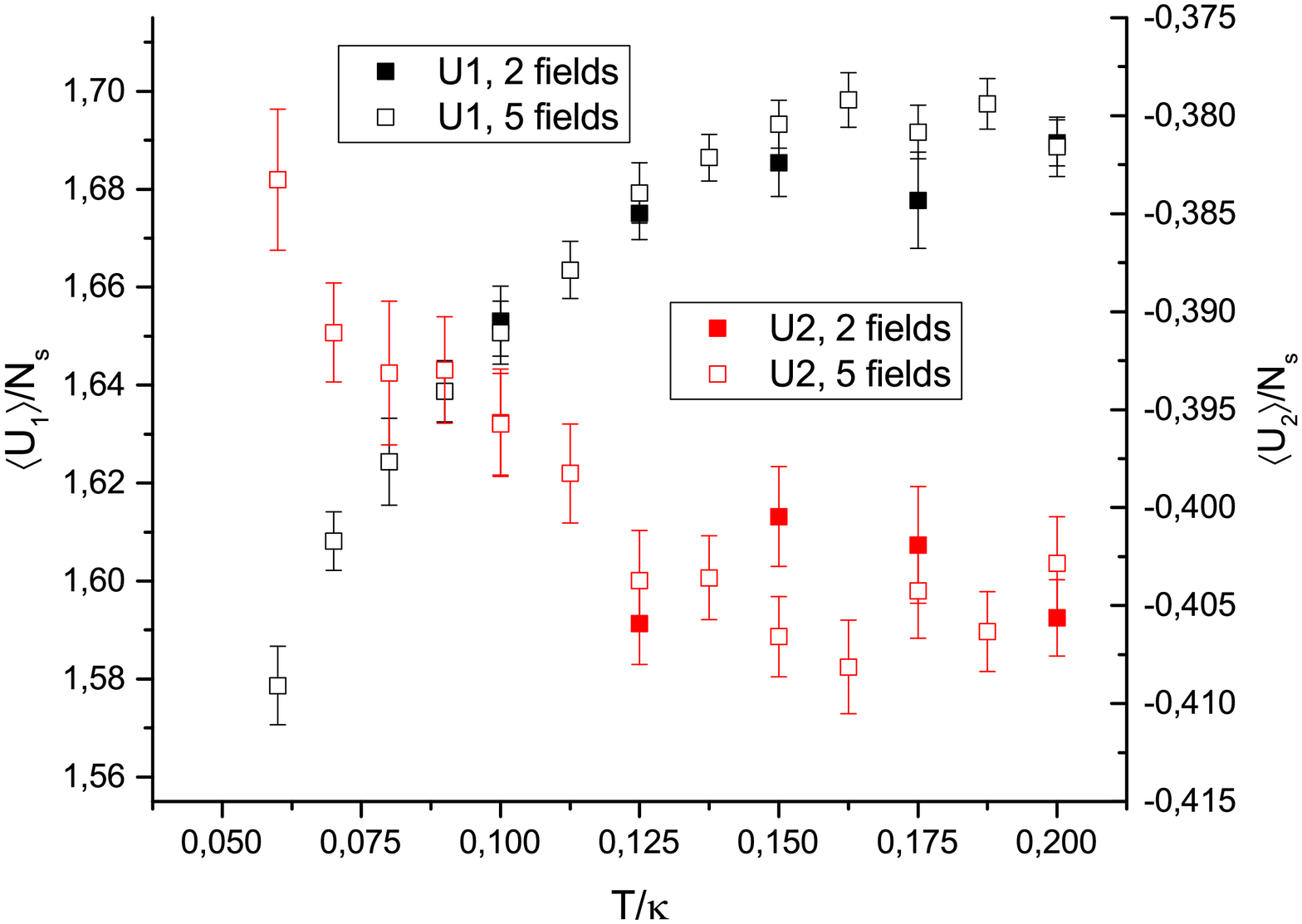}
		\caption{Two components of the potential energy $\langle \hat{U} \rangle$ per site in $6 \times 6$ lattice for low temperatures. While results for 2-fields fluctuate substantially, 5-fields approach stabilizes the computations so that a smooth curve can be traced within statistical error bars.}
		\label{fig:U1U2_cmp}
	\end{figure}

	To demonstrate how the modification affects the simulation results, let us look closely at the potential energy (\figurename~\ref{fig:pot_cmp}) and its components (\figurename~\ref{fig:U1U2_cmp}) at low temperatures. The values obtained in the 5-fields approach seem to be located smoother so that one can connect them with a continuous curve without violating the restrictions of statistical error bars. On the contrary, the results for 2 fields reveal an indication of metastable states, so that it requires several independent runs of a program to estimate the mean and standard deviation reasonably. Although the results do follow the same behaviour, the stability of the numerical results is essential in connection with a forthcoming calculation of derivatives of the (numerical) function or differences of large values. This takes place in the context of a heat capacity (defined both by a derivative and via energy variance), for example. So, due to the fact, that 5 fields help walking through a configuration space in a more flexible and gentle manner, which was supported by the data presented above, the distributions are modified with additional contributions from wide regions of a configuration space which make them more reliable for sequential calculations.
	
\begin{figure}
		\centering
		\includegraphics[width=0.9\linewidth]{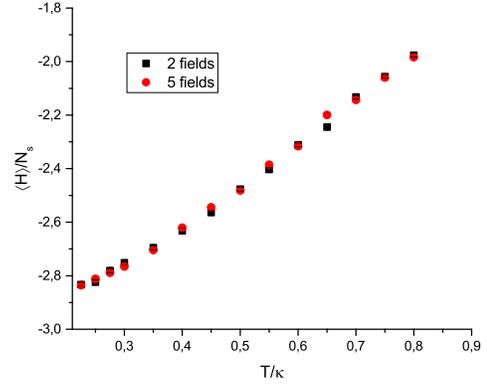}
		\caption{Thermal behaviour of $\langle \hat{H} \rangle$ (per site) for $6 \times 6$ in launches with 2 and 5 fields. The results are consistent.}
		\label{fig:H_cmp}
	\end{figure}

	If $\langle \hat{H} \rangle = \langle \hat{T} \rangle + \langle \hat{U} \rangle$ is computed, the overall result can be outlined as follows (\figurename~\ref{fig:H_cmp}). A regular kinetic part (with high agreement between 2-fields and 5-fields approaches) is superposed with a fluctuating potential part (with some deviations between the two cases). Actually, this \textit{can influence} the heat capacity determined by a derivative ($\partial \mathcal{E}/\partial T$, $\mathcal{E} = \langle \hat{H} \rangle$) for example, but appropriate fitting of the $\mathcal{E}(T)$ curve serves as a good remedy. On the other hand, as it was stated above, 5-field launches seem to produce more stabilized values, so the fitting may have better quality.

\begin{figure}
		\centering
		\includegraphics[width=0.9\linewidth]{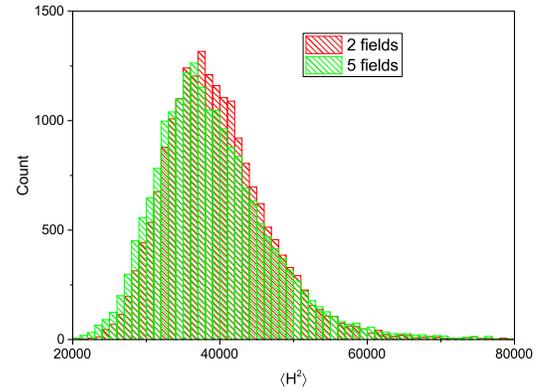}
		\caption{A histogram of $\langle \hat{H}^2 \rangle$. An analysis showed that an excess of non-exponential tail above the Gaussian fitting obtained from a good (left) part is reduced from 30.57\% to 28.78\% when involving 5 fields.}
		\label{fig:H2_hist}
	\end{figure}

\begin{figure}
		\centering
		\includegraphics[width=0.9\linewidth]{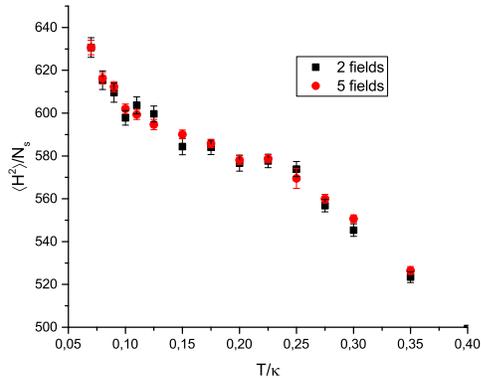}
		\caption{$\langle \hat{H}^2 \rangle$ (per site) for lattice $6\times 6$ in the cases of 2 and 5 fields applied. One can see a stable trend of a curve in the latter case. It improves a quality of fittings and assist processing the results when using $\langle \hat{H}^2 \rangle$ as a part of a complex mathematical expression. This is a part of \figurename~\ref{fig:H2_temp} for low temperatures.}
		\label{fig:H2_cmp}
	\end{figure}

	It is time to check $\langle \hat{H}^2 \rangle$. Let us do this by plotting a histogram of $\langle \hat{H}^2 \rangle$ values for 2-field and 5-field simulations (\figurename~\ref{fig:H2_hist}). It is noticeable that in the latter case a distribution is more regular at its top and has smoother side falls. There are more configurations with values of $\langle \hat{H}^2 \rangle \approx 22000 \mbox{ }(\mbox{eV}^2)$. It should be noted that these features do not vanish when temperature rises which is the case for $\langle \hat{T} \rangle$ and $\langle \hat{U} \rangle$, so the effect remains stable. This shows the difference between 4 and 8 or 6 operators multiplied. The distributions were tested at $T = 0.3\kappa$ and $T = 0.6\kappa$. An effect of a distribution refinement on the final results of $\langle \hat{H}^2 \rangle$ is depicted in \figurename~\ref{fig:H2_cmp}. Points corresponding to 5-fields launch fluctuate less intensively and the change of $\langle \hat{H}^2 \rangle (T)$ around $T=0.2\kappa$ is more apparent.

	It is proposed to put this \textit{mathematical} observations into practice by applying to heat capacity computation. As we have run into a number of nontrivial features while investigating $C$ as a function of temperature which are caused by complicated rearrangements in electron spacial density. For example, two phase transitions at $T = 0.028\kappa$ and $0.075\kappa$ are guessed with $V_{00} = 3.2\kappa$ and $V_{01} = 0.8\kappa$. Presumably they are related to formation and melting of an exciton condensate. The particular results obtained will be a topic of a separate paper.

	\section{Conclusion}
	
	The aim of our work was to generalize the method of Hubbard fields in fermion Monte Carlo simulation to the case of link fields. Such an improvement looks self-consistent from a physical point of view. The introduced Hubbard link fields play a role of the interaction fields responsible for the attraction and repulsion of electronic excitations at the nodes. In the paper we aim to show that the extension of the configuration space also helps hybrid Monte Carlo simulations due to improving the complex landscape of the extended Hubbard model that makes it difficult to bypass domain walls.
	
	Collecting the advantage which link Hubbard fields yield, one can state:
	\begin{enumerate}
		\item Thermalization time is reduces by several times;
		\item Autocorrelation time is also reduced especially in the region of low temperatures, where physical results are attractive;
		\item Fluctuation properties of observables are refined;
		\item Metastable states are depressed, they occur less frequently;
		\item An analytical form of action to compute using a computer is simplified, moreover existance of the boundary \(V_{01}=U_{00}/3\) becomes explicit;
		\item Heavy-tails of distributions represent a smaller contribution to the whole probability.
	\end{enumerate}
	
	In the work, we calculated the principle ingredients of the heat capacity of the gas of electron excitations in the extended Hubbard model on a hexagonal lattice. The results for $C$ itself and extensive analysis of its behaviour in vast range of temperatures will be presented in the next paper.
	
	It worth saying that the development proposed does not solve the problem of domain walls completely, but it helps Monte Carlo to observe a configuration space in a better way in comparison to the standard case. The statement can be proved by histogram examination which provides evidence for better distribution quality. Autocorrelation times are reduced and metastable states seem to emerge less often while using 5-fields approach.
	
	\begin{acknowledgements}
	
	
	The work was supported  by  grant  from  the  Russian  Science  Foundation  (project  number  21-12-00237).
	The research is carried out using the equipment of the shared research facilities of HPC computing resources at Lomonosov Moscow State University \cite{lomonosov} and Institute for Nuclear Research
	of the Russian Academy of Sciences.
	
	We would like to express our deepest appreciation to Dr. M. Ulybyshev for valuable discussions and guiding ideas that made this study possible. We are very grateful to the Referee for the careful reading of the paper and express our thanks for valuable advice and recommendations, which made it possible to improve the presentation of the content of the paper.
	
	\end{acknowledgements}
	
	\appendix
	
	\section{Form of the energy squared mean}
	\label{appA}
	
	An exact expression of energy squared in terms of fermion matrix (\ref{eq:fermmatrix}):
	
	\begin{widetext}
	\begin{multline*}
	\langle \hat{H}^2 \rangle = 2 \Re\Biggl\{ \kappa^2 \sum_{x,\mu,z,\nu} \left(P(w,y,x,z)
	+\delta_{w,x} M_{yz}^{-1}
	+M_{yx}^{-1} M_{wz}^{-1*} \right) \\
	-\kappa \sum_{x,\mu,z,\nu'} V_{\nu'} \Bigg( 2 P(w,y,z,w,x,z)
	+\delta_{w,x} P(y,z,w,z)
	+2 \delta_{w,z} P(w,y,w,x)
	-2 P(y,z,x,z) M_{ww}^{-1*}
	+\delta_{x,z} P(w,y,w,x) \\
	-\delta_{w,x}\delta_{x,z} M_{yw}^{-1}
	+\delta_{x,z} M_{yx}^{-1} M_{ww}^{-1*}
	-2 P(w,y,w,x) M_{zz}^{-1*}
	+\delta_{w,x} M_{yw}^{-1} M_{zz}^{-1*}
	+2 M_{yx}^{-1} P^{*}(w,z,w,z)
	-2 \delta_{w,z} M_{yx}^{-1} M_{ww}^{-1*} \\
	+\delta_{w,y} P(w,z,x,z)
	+\delta_{y,z} P(w,y,w,x)
	-\delta_{w,y}\delta_{w,z} M_{wx}^{-1}
	+\delta_{y,z} M_{yx}^{-1} M_{ww}^{-1*}
	+\delta_{w,y} M_{zz}^{-1} M_{wx}^{-1*} \Bigg) \\
	+\sum_{x,\mu',z,\nu'} V_{\mu'} V_{\nu'} \Biggl( P(w,x,y,z,w,x,y,z)
	-\delta_{w,x} P(w,y,z,w,y,z)
	-\delta_{w,y} P(w,x,z,w,x,z)
	-\delta_{w,z} P(w,x,y,w,x,y)\\
	+P(x,y,z,x,y,z) M_{ww}^{-1*}
	-\delta_{x,z} P(w,x,y,w,x,y)
	-\delta_{w,x}\delta_{x,z} P(w,y,w,y)
	-\delta_{w,y}\delta_{x,z} P(w,x,w,x)\\
	+\delta_{x,z} P(x,y,x,y) M_{ww}^{-1*}
	-\delta_{y,z} P(w,x,y,w,x,y)
	-\delta_{w,x}\delta_{y,z} P(w,y,w,y)
	-\delta_{w,y}\delta_{y,z} P(w,x,w,x)\\
	+\delta_{y,z} P(x,y,x,y) M_{ww}^{-1*}
	+P(w,x,y,w,x,y) M_{zz}^{-1*}
	+\delta_{w,x} P(w,y,w,y) M_{zz}^{-1*}
	+\delta_{w,y} P(w,x,w,x) M_{zz}^{-1*}\\
	+P(x,y,x,y) P^{*}(w,z,w,z)
	-\delta_{w,z} P(x,y,x,y) M_{ww}^{-1*}
	-\delta_{x,y} P(w,x,z,w,x,z)
	-\delta_{w,x}\delta_{x,y} P(w,z,w,z)\\
	-\delta_{w,z}\delta_{x,y} P(w,x,w,x)
	+\delta_{x,y} P(x,z,x,z) M_{ww}^{-1*}
	-\delta_{x,y}\delta_{x,z} P(w,x,w,x)
	+\delta_{w,x}\delta_{x,y}\delta_{x,z} M_{ww}^{-1}\\
	-\delta_{x,y}\delta_{x,z} M_{xx}^{-1} M_{ww}^{-1*}
	+\delta_{x,y} P(w,x,w,x) M_{zz}^{-1*}
	-\delta_{w,x}\delta_{x,y} M_{ww}^{-1} M_{zz}^{-1*}
	-\delta_{x,y} M_{xx}^{-1} P^{*}(w,z,w,z)
	+\delta_{w,z}\delta_{x,y} M_{xx}^{-1} M_{ww}^{-1*}\\
	+P(w,x,z,w,x,z) M_{yy}^{-1}
	+\delta_{w,x} P(w,z,w,z) M_{yy}^{-1*}
	+\delta_{w,z} P(w,x,w,x) M_{yy}^{-1}
	+P(x,z,x,z) P^{*}(w,y,w,y)\\
	-\delta_{w,y} P(x,z,x,z) M_{ww}^{-1*}
	+\delta_{x,z} P(w,x,w,x) M_{yy}^{-1*}
	-\delta_{w,x}\delta_{x,z} M_{ww}^{-1} M_{yy}^{-1*}
	-\delta_{x,z} M_{xx}^{-1} P^{*}(w,y,w,y)
	+\delta_{w,y}\delta_{x,z} M_{xx}^{-1} M_{ww}^{-1*}\\
	+P(w,x,w,x) P^{*}(y,z,y,z)
	-\delta_{w,x} M_{ww}^{-1} P^{*}(y,z,y,z)
	+M_{xx}^{-1} P^{*}(w,y,z,w,y,z)
	+\delta_{w,y} M_{xx}^{-1} P^{*}(w,z,w,z)\\
	+\delta_{w,z} M_{xx}^{-1} P^{*}(w,y,w,y)
	-\delta_{y,z} P(w,x,w,x) M_{yy}^{-1}
	+\delta_{w,x}\delta_{y,z} M_{ww}^{-1} M_{yy}^{-1*}
	+\delta_{y,z} M_{xx}^{-1} P^{*}(w,y,w,y)
	-\delta_{w,y}\delta_{y,z} M_{xx}^{-1} M_{ww}^{-1*} \Biggr) \Biggr\},
	\end{multline*}
	\end{widetext}
	where $V_\mu$ equals to $V_{00}/2$ if $\mu=0$ (so $y$ is $x$) and $V_{01}/2$ if $\mu=1,2,3$ ($y$ is a neighbour of $x$), $\mu = \overline{1,3}$, $\mu' = \overline{0,3}$, $y$ is found from $x$ and $\mu$ and $w$ is found from $z$ and $\nu$. Pairings are defined as follows:
	\begin{widetext}
	\begin{equation*}
	P(x, y, z, w) = M_{yz}^{-1} M_{xw}^{-1} - M_{xz}^{-1} M_{yw}^{-1},
	\end{equation*}
	\begin{equation*}
	P(x, y, z, w, u, v) = M_{xv}^{-1} P(y,z,w,u) - M_{xu}^{-1} P(y,z,w,v) +	M_{xw}^{-1} P(y,z,u,v),
	\end{equation*}
	\begin{equation*}
	P(x, y, z, w, u, v, t, s) = M_{xs}^{-1} P(y,z,w,u,v,t) - M_{xt}^{-1} P(y,z,w,u,v,s) +
	M_{xv}^{-1} P(y,z,w,u,t,s) - M_{xu}^{-1} P(y,z,w,v,t,s).
	\end{equation*}
	\end{widetext}


\begin{thebibliography}{99}
\bibitem{PhysRev.71.622} P. R. Wallace. The band theory of graphite. Phys. Rev.,
71:622–634, May 1947.
\bibitem{katsnelson_2012} M. I. Katsnelson. Graphene: Carbon in Two Dimensions.
Cambridge University Press, 2012.
\bibitem{buividovich2012numerical} P.V. Buividovich, E.V. Luschevskaya, O.V. Pavlovsky,
M.I. Polikarpov, and M.V. Ulybyshev. Numerical study
of the conductivity of graphene monolayer within the
effective field theory approach. Physical Review B,
86(4):045107, 2012.
\bibitem{ulybyshev2013monte} M.V. Ulybyshev, P.V. Buividovich, M.I. Katsnelson, and
M.I. Polikarpov. Monte carlo study of the semimetalinsulator
phase transition in monolayer graphene with
a realistic interelectron interaction potential. Physical
Review Letters, 111(5):056801, 2013.
\bibitem{valgushev2013influence} S.N. Valgushev, E.V. Luschevskaya, O.V. Pavlovsky,
M. I. Polikarpov, and M.V. Ulybyshev. Influence of defects
on the conductivity of graphene within the effective
theory approach. JETP letters, 98(7):389–392, 2013.
\bibitem{ulybyshev2015magnetism} M.V. Ulybyshev and M.I. Katsnelson. Magnetism and
interaction-induced gap opening in graphene with vacancies
or hydrogen adatoms: Quantum monte carlo study.
Physical Review Letters, 114(24):246801, 2015.
\bibitem{braguta2013numerical} V. V. Braguta, S.N. Valgushev, O.V. Pavlovsky, M. I.
Polikarpov, and M. V. Ulybyshev. Numerical simulation
of graphene in a magnetic field within the effective field
theory. JETP letters, 97(9):517–519, 2013.
\bibitem{braguta2013numerical2} V. Braguta, M.N. Chernodub, K. Landsteiner, M.I. Polikarpov,
and M.V. Ulybyshev. Numerical evidence of the
axial magnetic effect. Physical Review D, 88(7):071501,
2013.
\bibitem{boyda2014numerical} D.L. Boyda, V.V. Braguta, S.N. Valgushev, M.I. Polikarpov,
and M.V. Ulybyshev. Numerical simulation of
graphene in an external magnetic field. Physical Review
B, 89(24):245404, 2014.
\bibitem{buividovich2017interelectron} P. Buividovich, D. Smith, M. Ulybyshev, and L. von
Smekal. Interelectron interactions and the rkky potential
between h adatoms in graphene. Physical Review B,
96(16):165411, 2017.
\bibitem{PhysRevD.102.054502} M. Korner, K. Langfeld, D. Smith, and L. von Smekal.
Density of states approach to the hexagonal hubbard
model at finite density. Phys. Rev. D, 102:054502, Sep
2020.
\bibitem{PhysRevD.101.014508} M. Ulybyshev, Ch. Winterowd, and S. Zafeiropoulos. Lefschetz
thimbles decomposition for the hubbard model on
the hexagonal lattice. Phys. Rev. D, 101:014508, Jan
2020.
\bibitem{PhysRevB.98.235129} P. Buividovich, D. Smith, M. Ulybyshev, and L. von
Smekal. Hybrid Monte Carlo study of competing order in
the extended fermionic Hubbard model on the hexagonal
lattice. Phys. Rev. B, 98:235129, Dec 2018.
\bibitem{BSS} R. Blankenbecler, D. J. Scalapino, and R. L. Sugar. Monte Carlo calculations of coupled boson-fermion systems. I. Phys. Rev. D 24, 2278, October 1981.
\bibitem{ModPhysLetB} M. Shaw and W. P. Su. Phase Separation Due to Nearest Neighbor Attractive Interactions in a Two-Dimensional Model. Mod. Phys. Letters B, v. 17, n. 16.
\bibitem{PhysRevE.106.025318} M. Ulybyshev and F. Assaad. Mitigating spikes in
fermion monte carlo methods by reshuffling measurements.
Phys. Rev. E, 106:025318, Aug 2022.
\bibitem{PhysRevB.89.195429} D. Smith and L. Von Smekal. Monte-Carlo simulation
of the tight-binding model of graphene with partially
screened Coulomb interactions. Physical Review B, 89,
03 2014.
\bibitem{Ulybyshev2017PathIR} M. Ulybyshev and S. Valgushev. Path integral representation
for the hubbard model with reduced number of
lefschetz thimbles. arXiv: Strongly Correlated Electrons,
2017.
\bibitem{PhysRevB.89.205128} Wei Wu and A.-M. S. Tremblay. Phase diagram and fermi
liquid properties of the extended hubbard model on the
honeycomb lattice. Phys. Rev. B, 89:205128, May 2014.
\bibitem{PhysRevB.99.205434} P. Buividovich, D. Smith, M. Ulybyshev, and L. von Smekal. Numerical evidence of conformal phase transition in graphene with long-range interactions. Phys. Rev. B, 99:205434, May 2019.
\bibitem{Phys.Rev.Lett.106} T. O. Wehling, E. Şaşıoğlu, C. Friedrich, A. I. Lichtenstein, M. I. Katsnelson, and S. Blügel. Strength of Effective Coulomb Interactions in Graphene and Graphite. Phys. Rev. Lett. 106, 236805, June 2011.
\bibitem{doi:10.1142/9789814417891_0003} W. Janke. Monte Carlo Simulations in Statistical Physics
— From Basic Principles to Advanced Applications,
pages 93–166.
\bibitem{lomonosov} V. Sadovnichy, A. Tikhonravov, Vl. Voevodin, and
V. Opanasenko. ”Lomonosov”: Supercomputing at
Moscow State University, pages 283–307. Chapman \&
Hall/CRC Computational Science, Boca Raton, USA,
2013.
\end{thebibliography}

\end{document}